Heat capacity peak in solid $^4$He: Effects of disorder and $^3$He impurities

X. Lin, A. C. Clark*, Z. G. Cheng and M. H. W. Chan

Department of Physics, The Pennsylvania State University, University Park, Pennsylvania 16802, USA

Heat capacity measurements with significantly improved resolution find the presence of a peak in a solid $^4$He sample in coexistence with liquid. With improved crystallinity, the peak decreases in height and moves to lower temperature. A hysteretic heat capacity signature consistent with $^3$He-$^4$He phase separation, not detected in an earlier work is clearly observed.

Evidence of nonclassical rotational inertia (NCRI) in bulk solid helium first reported in 2004 [1, 2] has been confirmed in at least six other laboratories using the same torsional oscillator (TO) technique [3-8]. Although the temperature dependence of NCRI is qualitatively reproducible, the NCRI fraction (NCRIF) in these studies differs by three orders of magnitude, ranging from 0.03% to ~20%. As most TO experiments were carried out on polycrystalline samples grown by blocking the inlet capillary (BC) and freezing the high pressure liquid inside the torsion cell, it is generally believed that the variation in NCRIF is a consequence of disorder. Nevertheless, NCRI was found [9] in samples of large single crystals [10-12] grown by anchoring the TO cell at a constant pressure (CP) and temperature on the liquid-solid coexisting curve. However, differences do exist. For example, CP samples were found to possess a reproducibly sharp onset, and lower characteristic temperatures than those of BC samples grown in the same TO [9]. For the most thorough investigation discussed in Ref. 9, the overall temperature dependence in CP samples could be reproduced in BC samples only after ~25 h of annealing, strongly indicating the influence of sample quality. Moreover, quantum Monte Carlo simulations indicate that there should be no NCRI in a perfect crystal [13, 14]. An ideal system to test this prediction is a solid sample in coexistence with liquid, where defects such as grain boundaries and dislocations can be more readily annealed away [15-17]. Such a system has not been investigated with a TO.

In an earlier paper [18] we reported evidence of a heat capacity peak centering near 75 mK that rides on top of the Debye phonon contribution in solid $^4$He. The peak (both magnitude and temperature) appears to be independent of $^3$He concentration ($x_3$) for solid samples ranging from 1 ppb (isotopically pure $^4$He) to 10 ppm $^3$He, with the temperature of the peak coinciding with the onset of NCRI in CP samples of low $x_3$ [9]. The resolution of the measurements however, was limited such that the experimental uncertainty is nearly one fourth of the peak height of $2.5 \times 10^{-6}\,k_B$ per $^4$He atom. For samples with $x_3$ = 10 and 30 ppm, in addition to the heat capacity peak, there is an apparent temperature independent term in the heat capacity that increases in magnitude roughly proportionally with $x_3$. This constant term was interpreted in Ref. 18 to be related to the impurity ($^3$He) band reported in NMR measurements [19].

This interpretation however, is inconsistent with the theoretical expectation that such an impurity band should give rise to a heat capacity term that scales with $T^{-2}$ [19]. All solid samples studied in Ref. 18 were grown with the BC method over a time interval of $20\pm1$ h.

In the present work we investigate further the role of $^3$He, and for the first time focus on the dependence of the excess specific heat on various growth conditions. Measurements were performed in a new silicon calorimeter with nearly identical dimensions and design as that used in Ref. 18. In the first set of experiments 9 solid samples of different $x_3$, all grown with the BC method in 4 h, were studied. The protocol for the preparation of isotopic mixtures was similar to that in our earlier work [18]. Shortening the growth time from 20 to 4 h was accomplished by changing the thermal link between the calorimeter and the thermal bath. This also improved the minimum achievable temperature of the calorimeter from 40 to 25 mK. In a separate cool down, we thermally anchored the capillary such that we were able to grow a 1 ppb solid sample under the CP condition. The quoted pressure for each solid sample refers to the breakaway point from the melting curve that marks the time of growth completion. Lastly, we intentionally grew a solid-liquid coexistence sample with 75% solid and 25% liquid in the sample cell with the inlet capillary blocked with a solid $^4$He plug. For this sample $x_3 = 0.3$ ppm (commercially available ultrahigh purity $^4$He). Our measurement resolution at low temperature was improved by a factor of 4 with the use of a particularly sensitive (germanium chip) thermometer.

In Fig.1 the heat capacity $C_V$ at constant volume of samples prepared with different growth conditions are shown. The background heat capacity of the calorimeter, approximately 10 times smaller than that of solid $^4$He, has been subtracted. Deviations from the Debye behavior are found for all samples at low temperature. The heat capacity of 4 h BC samples with $x_3$ = 1 ppb and 0.3 ppm are identical within the scatter, as previously observed [18]. In panel (b) $C_V$ of a sample grown under CP is shown. A much smaller deviation from the Debye behavior is found. Two different samples are presented in panel (c): one of solid-liquid coexistence and one of only liquid. The magnitude of the latter is consistent to within 2% of that reported by Greywall [20]. Due to the heat leak along the filling capillary, we were unable to extend the measurements below 100 mK. Since the heat capacity due to liquid-solid conversion is negligible (~2 × 10$^{-10}$ J/K at 100 mK) [21], $C_V$ of the coexistence sample can simply be considered as the sum of the liquid and solid contributions. By comparing the known and measured values of $C_V$ for all liquid and all solid samples at temperatures above 150 mK, we deduce that the coexistence sample consists of 75% solid and 25% liquid by volume. The heat capacity of liquid $^4$He follows a $T^3$ dependence down to at least 65 mK [20, 22], which overlaps the temperature range of interest. Thus, the deviation found between 100 and 65 mK in panel (c) must have its origin in the solid. In addition, it is not unreasonable to assume that the entire heat capacity in excess to phonons is due to the solid phase since the anomaly would otherwise infer an order of magnitude increase in $C_V$ of the liquid between just 65 and

30 mK.

When the $T^3$ term extrapolated from high temperature (solid lines in each panel of Fig.1) is subtracted, a peak is found (see Fig. 2). The fact that the anomaly tends toward zero in the $T = 0$ limit supports the description of the data by a sum of two independent terms. The largest peak is found in BC samples grown in 4 h, and is about 48 μJ mol$^{-1}$ K$^{-1}$ (~5.8 × 10$^{-6}$ k$_B$ per $^4$He atom), followed by BC samples grown in 20 h at 20 μJ mol$^{-1}$ K$^{-1}$ [18] and then the CP sample at only 5 μJ mol$^{-1}$ K$^{-1}$. The temperature at which the peak occurs is also reduced. According to Refs. 9-12 and references therein, the density of defects in these samples is expected to follow the same trend. That is, Fig. 2 suggests that the excess specific heat decreases in magnitude and exists only at the lowest temperatures in samples of higher crystalline quality. This is consistent with NCRI findings and supports the notion that disorder, such as a dislocation network, is responsible for the observed phenomena. However, the independence of the heat capacity peak on $x_3$ for the two BC samples grown under identical conditions suggests that this is not simply due to $^3$He evaporation from dislocation lines.

The fact that a specific heat peak is found even for solid $^4$He in coexistence with liquid enlivens the discussion. Solid in coexistence with liquid can freely expand or contract, thus providing the best possible condition for relieving strain and annealing away defects. Melting curve pressure measurements cite a dislocation density limit of

$10^2$ cm$^{-2}$ (grown with an overpressure of 1 μbar) [15-17] for solid in coexistence of liquid, while the dislocation density in solid crystals usually ranges from $10^6$ to $10^9$ cm$^{-2}$ [11, 23, 24]. The actual dislocation densities in our samples are not known, but the lower peak temperature (55 versus 75 mK) and smaller excess entropy (2.5 versus 5.8 μJ mol$^{-1}$ K$^{-1}$) for the coexistence sample suggests that it is of higher quality than even the CP sample. This is not surprising since internal strains leftover from the crystal growth process should more easily relax in the presence of liquid. If the specific heat peak is indeed correlated with the onset of NCRI, then the finite anomaly in both the coexistence and CP samples support the notion that NCRI is intrinsic to solid helium and defects such as dislocation lines serve only to enhance the magnitude of the phenomenon [25].

The melting curve pressure of $^4$He was recently measured between 10 and 450 mK with a precision of approximately ±0.5 μbar [15-17]. In such a measurement, the entropy of the solid can be estimated from the pressure along the melting curve by assuming the low temperature entropy of the liquid is solely due to phonons. After correction of an instrumental effect, no additional entropy was found to exist in excess to phonons below 320 mK, with a stated limit six times smaller than that directly calculated from our coexistence sample. However, we note that $C_V/T$ is a direct measure of the derivative of entropy and is inherently more sensitive and less unambiguous than using the melting curve to detect any anomaly. For instance, any pressure anomaly, no matter how small, will result in an arbitrarily large excess entropy if it occurs over an arbitrarily small temperature range. On the other hand, an experimentally unresolved change in pressure of 0.5 μbar, if uniformly spread over 80 mK (the approximate width of the heat capacity peak) would correspond to an excess entropy four times larger than the limit set by Refs. 15-17. In contrast, the entropy and also the pressure can be directly calculated by integrating $C_V/T$. We arrive at a deviation in the pressure of <1 μbar from Debye behavior for the specific heat measured on the melting curve.

We have studied seven additional samples with higher $^3$He concentrations to elucidate the puzzle of $^3$He-$^4$He mixtures reported in Ref. 18. These were BC samples grown in 4 h. Due to the heat flush effect induced by the thermal gradient between the cell and the capillary, and the adsorption of $^3$He on the liquid-solid interface [26-28], the concentration of a solid inside the calorimeter is often considerably (and not predictably) lower than that of the starting gaseous mixture. For simplicity, these mixture samples are still identified by their gaseous concentrations: 5, 10(a), 10(b), 30, 100(a), 100(b), and 500 ppm. In Fig. 3, samples 10(a) and 500 ppm show a deviation from a $T^3$ dependence at significantly higher temperature than that of the 1 ppb and 0.3 ppm samples. The most likely explanation is $^3$He-$^4$He phase separation (PS). Hysteresis and long equilibration times, characteristic signatures of PS, were observed in all seven samples with $x_3 \geq 5$ ppm. This behavior was also observed previously in our laboratory in a sample with a large $^3$He concentration of $x_3 = 760$ ppm [29]. In Ref. 18 this behavior may have been overlooked in dilute samples ($x_3 = 10$ and 30 ppm) due to insufficient resolution to detect small drifts of $C_V$ in time, resulting in an apparent temperature independent term. The inset of Fig. 3 shows the nonphonon heat capacity of 1 ppb, 0.3, 10(a), 30, 100(b) and 500 ppm samples measured during a cooling scan. For $\mathbf{x}_3$ = 1 ppb and 0.3 ppm, only one peak is observed, as shown in Fig. 2. For higher $x_3$, the heat capacity data hug the peak found in dilute mixtures until upturning at a lower temperature. These upturns occur for samples with $x_3$ = 10(a), 10(b), 30, 100(a), 100(b) and 500 ppm at 63, 74, 76, 66, 86 and 109 mK respectively, with an uncertainty of $\pm 5$ mK. If treated as PS temperatures, according to the model of Edwards and Balibar [30], the extracted $x_3$ values of these samples are 0.8, 7.5, 9, 1.5, 36 and 290 ppm, respectively, fairly consistent with our expectations. The $^3$He anomaly of the 5 ppm sample is too small to determine the upturn temperature.

The most striking result revealed by the inset of Fig. 3 is that, prior to PS, the excess specific heat of a sample with $x_3$ as high as 290 ppm is almost identical to that of a sample with just 1 ppb $^3$He. This result is in strong contrast to the findings in TO measurements where the onset of NCRI was dramatically broadened from 75 to 280 mK by the addition of 10 ppm $^3$He [24]. Anderson describes NCRI as the rotational susceptibility of a vortex liquid phase [31]. The true transition for the vortex state independent of $^3$He impurities is posited to occur at a temperature much lower than the onset temperature of NCRI. A frequency effect (higher onset temperature for higher frequency at fixed $x_3$) qualitatively consistent with the model of Anderson was observed in a TO with two resonant frequencies [4]. It was recently found that in a sample of higher $x_3$, the frequency dependence of the onset point is even more dramatic [32]. These TO experiments together with the current result suggest that the $^3$He dependence in NCRI may only be a dynamical effect.

In conclusion, evidence for isotopic phase separation was found in samples containing as little as 10 ppm $^3$He. Independently, a specific heat peak has been reproduced from earlier measurements. In contrast to NCRI, the peak is independent of $^3$He impurities. However, it does exhibit a dependence on sample quality that is very similar to the findings on NCRI. The peak remained present in a sample under even the lowest possible strain, solid in coexistence with liquid. This supports the interpretation that the specific heat peak is the thermodynamic signature of a supersolid state.

We thank J. T. West and P. W. Anderson for discussions and J. A. Lipa for providing us with isotopically pure $^4$He. This work is supported by NSF under grant DMR-0706339.

*Present address: Department of Physics, University of Basel, CH-4056 Basel, Switzerland.

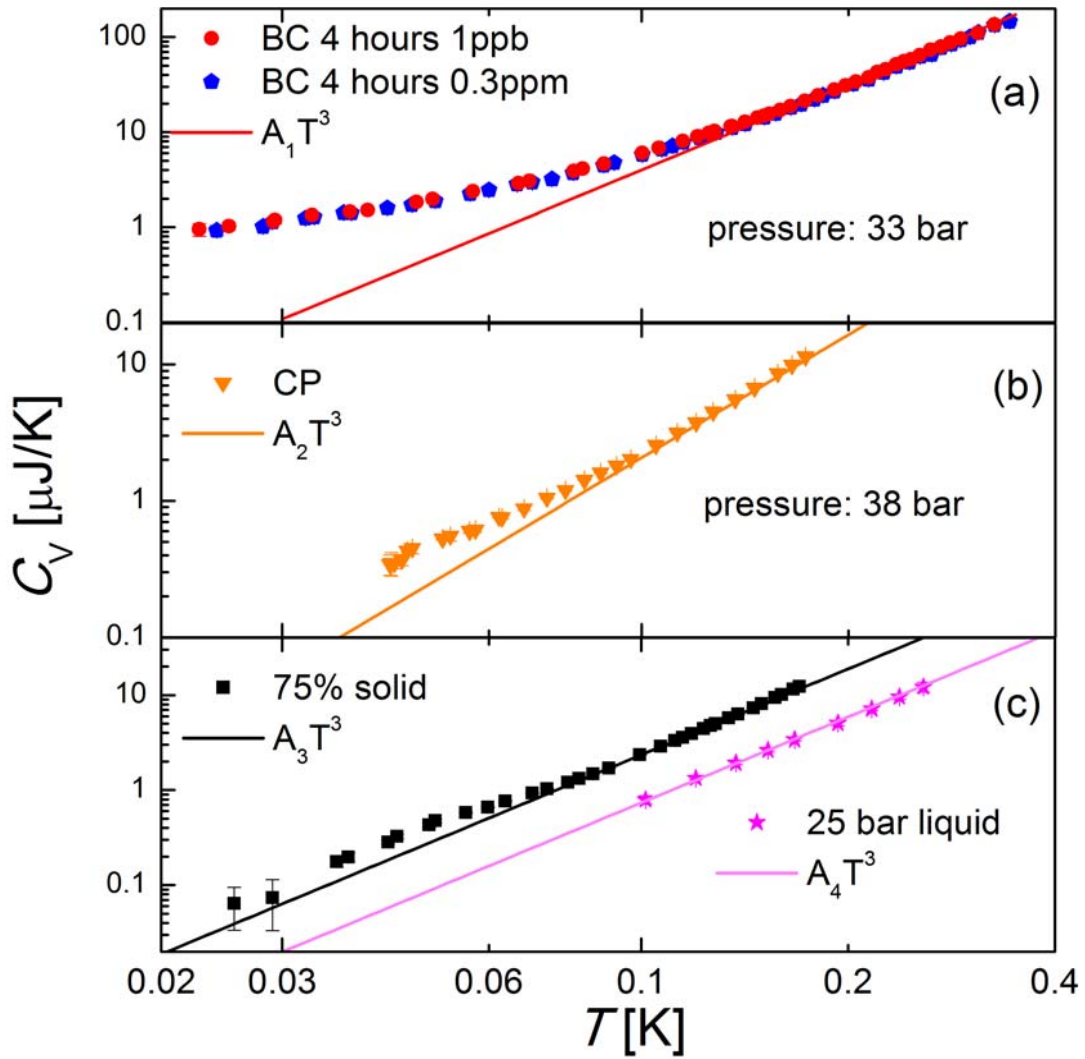

Fig. 1: From top to bottom, the heat capacity data of (a) BC $^4$He samples (4 h growth time) with $x_3 = 1$ ppb (red circles) and 0.3 ppm (blue pentagons), (b) constant pressure sample with $x_3 = 1$ ppb (orange triangles), (c) solid-liquid coexistence sample with $x_3 = 0.3$ ppm (black squares) and a 25 bar liquid sample with $x_3 = 1$ ppb (pink stars). The densities of the three solid samples are 20.46, 20.46 and 20.20 cm$^3$ mol$^{-1}$, respectively. The scatter of the data, unless shown otherwise, is less than or equal to the size of the symbols. The volume of the calorimeter is 0.93 cm$^3$.

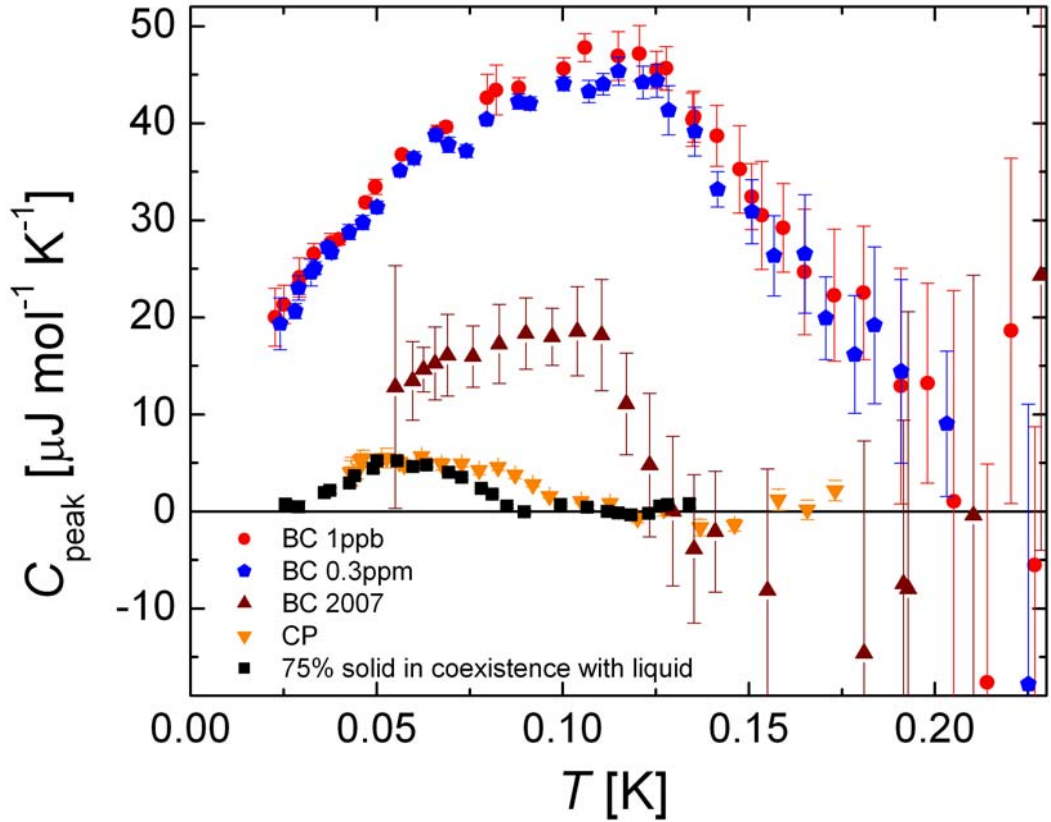

Fig. 2: Specific heat peak as a function of temperature for solid $^4$He samples grown under different conditions. From top to bottom: (1) BC $^4$He samples, 4 h growth time with $x_3$ = 1 ppb (red circles) and $x_3$ = 0.3 ppm (blue pentagons), (2) BC $^4$He sample, 20 h growth time with $x_3$ = 1 ppb (brown triangles) [18], (3) CP sample with $x_3$ = 1 ppb (orange triangles), (4) solid sample, 75% by volume in coexistence with liquid, with $x_3$ = 0.3 ppm (black squares). The data at higher temperature have larger scatter because the uncertainty scales with the total measured $C_V$.

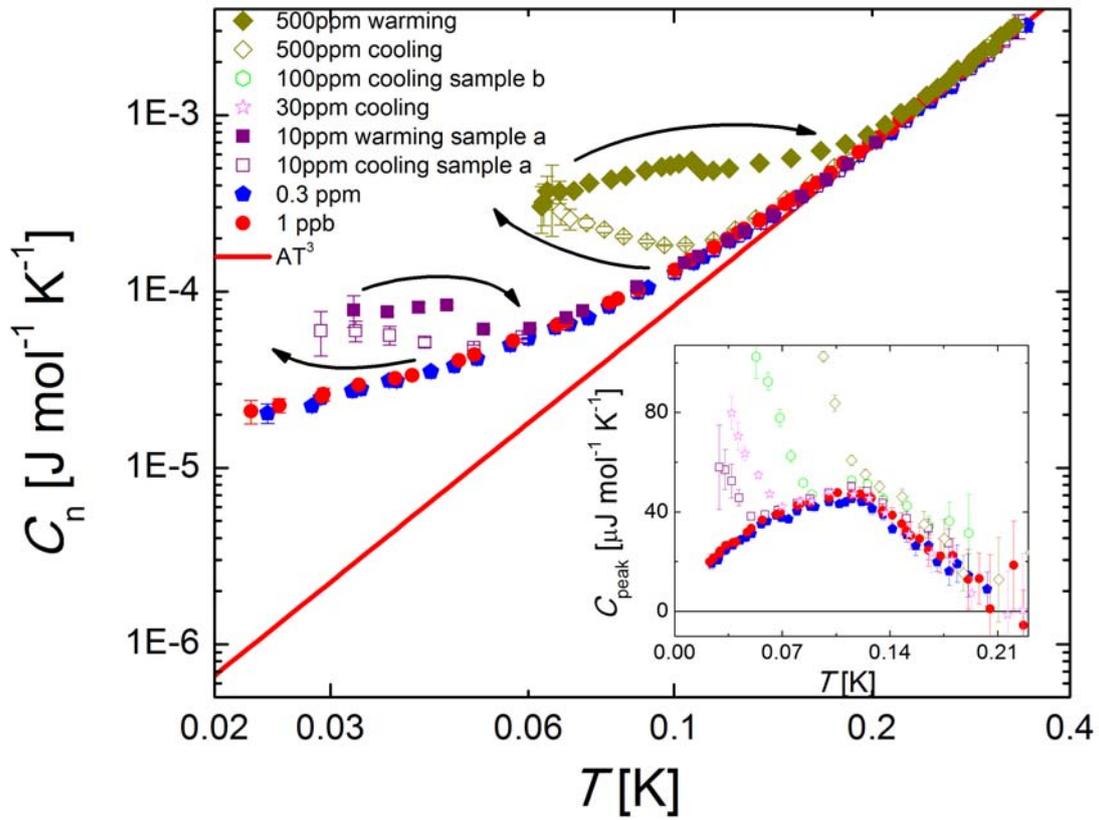

Fig. 3: Hysteresis in the molar specific heat of samples with $x_3$ = 10(a) and 500 ppm, which we attribute to isotopic PS. 1 ppb and 0.3 ppm samples of similar density ($P$ = 33±1 bar) are shown together. The inset includes the specific heat in excess to the phonon contribution of BC samples with $x_3$ = 1 ppb, 0.3, 10(a), 30, 100(b) and 500 ppm, obtained in cooling scans.